\documentclass[12pt]{article}

\usepackage{amssymb,hyperref}

\def\nqq{\hspace{-2em}}

\def\barr{\left(\begin{array}}
\def\earr{\end{array}\right)}
\def\beq#1{\begin{equation}\label{#1}}
\def\eeq{\end{equation}}
\def\ber#1{\begin{eqnarray}\label{#1} &&\nqq}
\def\eer{\end{eqnarray}}

\newcommand{\bear}[1]{\begin{eqnarray}\label{#1}}
\newcommand{\ear}{\end{eqnarray}}

\renewcommand{\theequation}{\arabic{section}.\arabic{equation}}
\catcode`\@=11 \@addtoreset{equation}{section}\catcode`\@=12

\newcommand{\R}{ {\mathbb R} }

\newcommand{\fnm}{\footnotemark}
\newcommand{\fnt}{\footnotetext}

\begin{document}

 \begin{center}

  \large \bf
 Exact solutions  in  gravity with a
  sigma model source

  \end{center}

\vspace{15pt}

\begin{center}

 \normalsize\bf
        A. A. Golubtsova\fnm[1]\fnt[1]{siedhe@gmail.com}$^{, b, c}$
        and   V. D. Ivashchuk\fnm[2]\fnt[2]{ivashchuk@mail.ru}$^{, a, b}$

 \vspace{7pt}

 \it (a) \ \ \ Center for Gravitation and Fundamental
 Metrology,  VNIIMS, 46 Ozyornaya Str., Moscow 119361, Russia  \\

 (b) \  Institute of Gravitation and Cosmology,
 Peoples' Friendship University of Russia,
 6 Miklukho-Maklaya Str.,  Moscow 117198, Russia \\

(c) \  Laboratoire de Univers et Th\'{e}ories (LUTh), Observatoire de Paris,\\
Place Jules Janssen 5, 92190 Meudon, France \\

 \end{center}
 \vspace{15pt}



\begin{abstract}
We consider a  $D$-dimensional model of gravity with non-linear
``scalar fields'' as a matter source. The model is defined on the
product manifold $M$, which contains $n$ Einstein factor spaces.
General cosmological type solutions  to the field equations are
obtained when $n-1$ factor spaces are Ricci-flat, e.g. when one
space $M_1$ of dimension $d_1 > 1$ has nonzero scalar curvature.
The solutions are defined up to solutions to geodesic equations
corresponding to a sigma model target space. Several examples of
sigma models are presented.  A subclass of spherically symmetric
solutions is studied and a restricted version of ``no-hair
theorem'' for  black holes is proved. For the case $d_1 =2$ a
subclass of latent soliton solutions   is singled out.

\end{abstract}

\section{Introduction}

A lot of gravitational models (including scalar-tensor theories
and supergravity ones), when required solutions have enough
space-time symmetries, can be reduced  to an effective  model with
``scalar fields'' governed by a sigma model action, see \cite{BMG,
GK, IMC, Chervon, GR, BM,Cl, CFRSTVR}  and references therein.

Here we consider a gravitational model  governed  by a Lagrangian
in $D$ dimensions

 \beq{1.1}
 {\cal L} = R[g] - h_{a
b}(\varphi)g^{MN}\partial_{M}\varphi^{a}\partial_{N}\varphi^{b},
\eeq

 where $g$ is a metric and non-linear ``scalar fields''
 $\varphi^\alpha$ come to the Lagrangian in a sigma model
form  with a certain target space metric $h$ assumed. In higher
dimensional gravitational theories the above model first appeared
in connection with spontaneous compactification of the extra
dimensions \cite{OP,GMZ}.

 The Lagrangian (\ref{1.1}) (e.g. with $h_{a b} = const$) describes the
truncated  sectors of various $3 < D \leq 10$ supergravity
theories in the Einstein frame  \cite{SaSe}. Usually these
theories contain form-fields (fluxes) in addition to massless
scalar fields and Chern-Simons (CS) terms. In this sense, the
Lagrangian (\ref{1.1})  for $D > 3$ (e.g. with $h_{a b} = const$)
matches zero flux (and CS) limit. For $D =3$ the Lagrangians of
such types are generic ones when dimensional reductions of
(bosonic sectors of) supergravity models are considered, see
\cite{BMG, GK, BM, Cl} and refs. therein. In higher dimensional
gravitational theories the above model first appeared in
connection with spontaneous compactification of extra dimensions
\cite{OP,GMZ}.

Here we deal with cosmological type solutions defined on a product
of $n$ Einstein spaces (e.g. Ricci-flat ones). The integrable
cosmological configurations were studied in numerous papers, see
\cite{I-92}, \cite{GIM} (without scalar fields),
\cite{BZ0,BZ1,BZ2,IM10} (with one scalar field),
\cite{I-03,I-05,BZhuk} etc. However, the authors of these papers
restricted their attention to a linear sigma model for which
components $h_{ab}$ are constant. Here we will study the solutions
for $h_{a b}(\varphi)$ with arbitrary dependence on $\varphi^{a}$
(e.g. for a non-linear sigma model).

\section{The model}

We start by considering the following action

\beq{2.1} S = \frac{1}{2\kappa^{2}}\int_{M} d^{D}x
\sqrt{|g|}\Bigl\{R[g] - h_{ab}(\varphi)
g^{MN}\partial_{M}\varphi^{a}\partial_{N}\varphi^{b}\Bigr\} +
S_{YGH}, \eeq

where  $\kappa^{2}$ is a $D$-dimensional gravitational coupling,
$g = g_{MN}dx^{M}\otimes dx^{N}$ is a metric defined on a manifold
$M$, $\varphi: M \rightarrow M_{\varphi}$ is a smooth sigma model
map and $M_{\varphi}$ is a $l$-dimensional manifold (target space)
equipped with the metric $h = h_{a b}(\varphi) d \varphi^a \otimes
d \varphi^b$ ($\varphi^a$ are coordinates on $M_{\varphi}$). Here
$S_{YGH}$ is the standard York-Gibbons-Hawking boundary term
\cite{Y,GH}.

The field equations for the action (\ref{2.1}) read as follows
 \bear{1.4E}
 R_{MN} - \frac12 g_{MN}R = T_{MN},\\
 \label{1.4F} \frac{1}{\sqrt{|g|}}\partial_{M}(g^{MN}\sqrt{|g|}
 h_{a b}(\varphi)\partial_{N}\varphi^{b}) - \frac{1}{2} \frac{\partial
 h_{b c}(\varphi)}{\partial \varphi^{a}}
 \partial_{K}\varphi^{b}\partial_{L} \varphi^{c} g^{KL} = 0,
\ear

where

 \bear{1.4T} T_{MN} =
 h_{a b}(\varphi)\partial_{M}\varphi^{a}\partial_{N}\varphi^{b}
 - \frac12 h_{a b}(\varphi)g_{MN}
 \partial_{K}\varphi^{a} \partial_{L} \varphi^{b} g^{KL}
 \ear

 is the stress-energy tensor.

Here we consider a cosmological type  ansatz  for the  metric and
``scalar fields''

 \bear{2.2} g =  w e^{2\gamma(u)} du\otimes du +
 \sum^{n}_{i=1}e^{2\beta^{i}(u)}g^{i}, \\ \label{2.2F}
 \varphi^{a}  = \varphi^{a}(u),
 \ear
 where $w = \pm 1$, $i = 1, \ldots, n$.

The metric is defined on the  manifold

 \beq{2.3} M = \mathbb{R}_{*}\times M_{1} \times\ldots \times
           M_{n}, \eeq

where $\mathbb{R}_{*} = (u_{-}, u_{+})$ and any  factor space
$M_{i}$ is a $d_{i}$-dimensional Einstein manifold with the metric
$g^{i}$ obeying
 \beq{Ein}
  R_{m_{i}n_{i}}[g^{i}] = \xi_{i} g^{i}_{m_{i}n_{i}},
 \eeq
 $i = 1, \dots, n$.

To find solutions for the equations (\ref{1.4E})-(\ref{1.4F})
seems to be complicated due to the non-linear structure of the
Einstein equations and intricacy having scalar fields. However it
may be shown that  the field equations for the model (\ref{2.1})
with the metric and ''scalar fields'' from (\ref{2.2}) and
(\ref{2.2F}) are equivalent to the Lagrange equations
corresponding to the Lagrangian of the one-dimensional
$(n+l)$-component modified $\sigma$-model

 \beq{3.L} L = \frac 12
 {\cal N}^{-1} [ G_{ij}\dot{\beta}^{i}\dot{\beta}^{j} +
 h_{a b}(\varphi)\dot{\varphi}^{a} \dot{\varphi}^{b} ]
   -  {\cal N} V_{\xi}.
 \eeq

 Here ${\cal N} =  \exp(\gamma - \gamma_{0}) > 0$ is a modified lapse
 function, $\gamma_{0} =  \displaystyle{\sum^{n}_{i=1}d_{i}\beta^{i}}$,

  \beq{3.3} G_{ij} = d_{i}\delta_{ij} - d_{i}d_{j},  \eeq

  $i,j = 1, \ldots,n$,   are components of the gravitational part of the
minisuperspace metric \cite{IMZ} and

\beq{3.2} V_{\xi} =
        \frac{w}{2}\sum^{n}_{i=1}\xi_{i}d_{i}e^{-2\beta^{i} + 2\gamma_0}
\eeq

is the potential. Here and in what follows $\dot{A} =
\frac{dA}{du}$.

For the constant $h_{a b}(\varphi) = h_{a b}$ the reduction to the
sigma model was proved (for more general setup) in \cite{IMC}. We
note that here $h_{a b}(\varphi)$ can be interpreted as a scalar
part of the total target space metric.

When all $M_i$ have finite volumes  the substitution of (\ref{2.2}) and
(\ref{2.2F}) into  the action (\ref{2.1}) gives us the following
relation

\beq{3.1} S = \mu \int L du  \eeq

where $L$ is defined in (\ref{3.L}),
 $\mu = -\displaystyle{\frac{w}{\kappa^{2}} \prod_{i =1}^n
 V_{i}}$, and $V_{i} = \displaystyle{\int_{M_{i}} d^{d_{i}}y
 \left(\sqrt{\det{(g^{i}_{m_{i}n_{i}})}}\right)}$ is the volume of
$M_{i}$, $i = 1, \ldots,n$.

The relation (\ref{3.1})  can be derived using the following
expression for the scalar curvature

\beq{3.ScCurv} R = -w e^{-2\gamma}\Bigl(2\ddot{\gamma}_{0} -
             2\dot{\gamma}_{0} \dot{\gamma} + \dot{\gamma}^{2}_{0} +
\sum^{n}_{i=1}d_{i}\left(\dot{\beta}^{i}\right)^{2}\Bigr) +
\sum^{n}_{i=1}e^{-2\beta^{i}}R[g^{i}], \eeq

where $R[g^{i}] = \xi_{i} d_i$ is the scalar curvature
corresponding to the $M_{i}$-manifold.

To obtain (\ref{3.1}) from (\ref{2.1}) (for compact $M_i$) one
should extract the total derivative term in  (\ref{3.ScCurv})
which is cancelled by the York-Gibbons-Hawking boundary term.

We write the Lagrange equations for (\ref{3.L}) and then put
${\cal N} = 1$, or equivalently $\gamma = \gamma_{0}$, i.e. when
$u$ is a harmonic variable, as in \cite{KA-73}. We get

 \beq{3.4}
 G_{ij}\ddot{\beta}^{j} +
 w \sum^{n}_{j=1}\xi_{j}d_{j} ( - \delta_i^j  + d_i) e^{-2\beta^{j} +
 2\gamma_0} = 0,
 \eeq
 $i = 1, \ldots,n$,
 \beq{3.5}
 \frac{d(h_{a b}(\varphi)\dot{\varphi}^{b})}{d u} -
 \frac12 \frac{\partial h_{c b}(\varphi)}{\partial
 \varphi^{a}} \dot{\varphi}^{c} \dot{\varphi}^{b} = 0,
 \eeq
 $a = 1,\ldots, l$,  and
\beq{3.6} \frac{1}{2}
  G_{ij}\dot{\beta}^{i}\dot{\beta}^{j} +   \frac{1}{2}
   h_{a b}(\varphi)\dot{\varphi}^{a}\dot{\varphi}^{b}
    + V_{\xi} = 0. \eeq

In fact,  equations (\ref{3.4}) are nothing else but Lagrange
equations corresponding to the Lagrangian
 \beq{3.10} L_{\beta} = \frac 12
    G_{ij}\dot{\beta}^{i}\dot{\beta}^{j} - V_{\xi}
 \eeq
with the energy integral of motion
 \beq{3.10e} E_{\beta} = \frac 12
    G_{ij}\dot{\beta}^{i}\dot{\beta}^{j} + V_{\xi}.
 \eeq

Likewise (\ref{3.4}), the equations (\ref{3.5}) are Lagrange equations corresponding to the
Lagrangian
\beq{3.11} L_{\varphi} = \frac 12
    h_{a b}(\varphi)\dot{\varphi}^{a}\dot{\varphi}^{b}
 \eeq
 with the energy integral of motion
\beq{3.11e}
  E_{\varphi} = \frac 12
  h_{a b}(\varphi)\dot{\varphi}^{a} \dot{\varphi}^{b}.
 \eeq
Equations (\ref{3.5}) are equivalent to geodesic equations
corresponding to the metric $h$.

The relation (\ref{3.6}) is the energy constraint

\beq{3.12} E = E_{\beta} + E_{\varphi} =0,
\eeq
coming from $\partial L/ \partial {\cal N} = 0$ (for ${\cal N }=  1$).

Equations  (\ref{3.4}) can be rewritten in a equivalent form

 \beq{3.4I}
 \ddot{\beta}^{i} -  w \xi_{i} e^{-2\beta^{i} +  2\gamma_0} = 0,
 \eeq
  $i = 1, \ldots,n$. These expressions may be obtained from (\ref{3.4})
  by using the inverse matrix $(G^{ij}) = (G_{ij})^{-1}$:
\beq{3.3I}
 G^{ij} = \frac{\delta^{ij}}{d_{i}} + \frac{1}{2 - D}
\eeq
 and the following relations for  $u^{(k)}$-vectors:
\beq{3.3u}
  u^{(k)}_i =  - \delta^k_i  + d_i,  \qquad  u^{(k)i} = G^{ij}
  u^{(k)}_j = - \frac{\delta^{ki}}{d_i},
 \eeq
 $i,j, k = 1, \ldots,n$.

In what follows we will use the following formulae
  \beq{3.13}
   (u^{(k)},u^{(k)}) = G^{ij} u^{(k)}_i u^{(k)}_j = \frac{1}{d_k}
   -1,
  \eeq
  $k = 1, \ldots,n$.

Hence the problem of finding the cosmological type solutions for
the model (\ref{2.1})
 (with   $u$ being  harmonic variable) is reduced
 to solving the equations of motion for the Lagrangians
 $L_{\beta}$ and $L_{\varphi}$ with the energy constraint (\ref{3.12}) imposed.

{\bf Geodesics for a flat metric $h$.}

For the constant $h_{a b}(\varphi) = h_{a b}$  eqs. (\ref{3.5})
read

 \beq{3.14} \ddot{\varphi}^{a} = 0,
 \eeq
 or, equivalently,

\beq{3.15} \varphi^{a} = \upsilon^{a}_{\varphi} u + \varphi^{a}_0,
\eeq

where $\upsilon^{a}_{\varphi}$ and $ \varphi^{a}_0$ are
integration constants, $a = 1, \dots, l$.

 The energy for scalar fields (\ref{3.11e}) takes the form

 \beq{4.12}
  E_{\varphi} = \frac 12 h_{ab} \upsilon^{a}_{\varphi}
 \upsilon^{b}_{\varphi}.
 \eeq

More examples of geodesic solutions will be given in Section 4.

\section{Cosmological type solutions}

In this section we deal with certain examples of cosmological type
solutions with the  metric and  "scalar fields'' from
 (\ref{2.2}) and  (\ref{2.2F}), respectively.

 \subsection{Solutions with $n$ Ricci-flat spaces}

Here  we focus on the solutions for the case when all factor
spaces $M_{i}$ are Ricci-flat:
 \beq{4.1}
  {\rm Ric}[g^{i}] = 0,
 \eeq
$i=1,\ldots,n$.

Due to (\ref{4.1}) the potential $V_{\xi}$ is equal to zero and
the equations of motion (\ref{3.4I}) for $\beta^{i}$ now become
\beq{4.2} \ddot{\beta}^{i} = 0, \eeq

$i=1,\ldots,n$.

Integration of the equations (\ref{4.2}) yields

\beq{4.3} \beta^{i} = \upsilon^{i} u + \beta^{i}_{0}, \qquad
\gamma_0 = \sum^{n}_{i=1}d_{i}\left(\upsilon^{i}u +
\beta^{i}_{0}\right), \eeq

where the parameters $\upsilon^{i}$ and $\beta^{i}_{0}$ are
integration constants and the energy  (\ref{3.10e}) takes the form

\beq{4.8} E_{\beta} = \frac{1}{2}G_{ij}v^{i}v^{j}, \eeq

where the minisuperspace metric $G_{ij}$ is given by (\ref{3.3}).

The  metric reads

 \beq{4.4} g = w \exp{[2\sum^{n}_{i=1}d_{i}(\upsilon^{i}u
 + \beta^{i}_{0})]} du \otimes du +
 \sum^{n}_{i=1}\exp{[2(\upsilon^{i}u + \beta^{i}_{0})]}g^{i}. \eeq

The ``scalar fields'' obey eqs. (\ref{3.5}) with the energy
constraint

\beq{4.6} E_{\varphi} = \frac 12
  h_{a b}(\varphi)\dot{\varphi}^{a} \dot{\varphi}^{b} =  -\frac{1}{2}G_{ij}v^{i}v^{j}. \eeq

In a special case of one (non-fantom) scalar field ($h_{11} = 1$)
and $w= -1$ this solution was obtained in \cite{BZ0}. For
solutions with several scalar fields and constant $h_{ab}$ see
\cite{I-03,I-05}.

The scalar curvature for the metric (\ref{4.4}) reads (see
        (\ref{3.ScCurv}))

\beq{4.7} R[g] = -w
\left(G_{ij}\upsilon^{i}\upsilon^{j}\right)e^{-2\gamma_{0}}. \eeq

In what follows we use a parameter

\beq{4.Sigma} \Sigma =\displaystyle{\Sigma(\upsilon)
              \equiv\sum^{n}_{i=1}d_{i}\upsilon^{i}} \eeq

to classify the solutions.\\

\textbf{Non-special Kasner-like solutions.}

 We shall first consider the non-special
case when $\Sigma(\upsilon) \neq 0$.

Let us define  a  ``synchronous'' variable

 \beq{4.tau} \tau =
  \frac{1}{|\Sigma(\upsilon)|}\exp{\Bigl[\sum^{n}_{j=1}(\upsilon^{j}
  u  + \beta^{j}_{0})d_{j}\Bigr]} \eeq

 obeying $e^{2 \gamma_0(\beta)}du^{2} = d \tau^2$.

 We introduce new  parameters:

  \beq{4.alpha} \alpha^{i} =   \upsilon^{i} / \Sigma(\upsilon),
  \eeq
  $i = 1,\ldots, n$, and

 \beq{4.Ea} \mathcal{E}_{\varphi} = E_{\varphi} / (\Sigma(\upsilon))^2.
 \eeq

 Then  the metric reads

 \beq{4.g} g = w d\tau \otimes \tau +
           \sum^{n}_{i=1}c^{2}_{i}\tau^{2\alpha_i} g^{i}, \eeq

 $\tau > 0$. ``Scalar fields'' are solutions to equations of motion   (see (\ref{3.5}))

   \beq{4.phi}
  \frac{d}{d \tau } \left[ \tau h_{a b} (\varphi) \frac{d \varphi^{b}}{d \tau} \right]  -
  \frac{1}{2} \tau \frac{\partial h_{c b}(\varphi)}{\partial
  \varphi^a} \frac{d \varphi^c}{d \tau} \frac{d \varphi^{b}}{d
  \tau} =  0,
  \eeq
   $a = 1, ..., l$.
 The parameters (\ref{4.alpha}) obey the Kasner-like  conditions

 \bear{4.Kas1} \sum^{n}_{i=1}d_{i}\alpha^{i} = 1, \\
      \label{4.Kas2}
  \sum^{n}_{i=1}d_{i}(\alpha^{i})^{2} = 1 - 2 \mathcal{E}_{\varphi},
\ear
  where

      \beq{4.Ec} \displaystyle{2 \mathcal{E}_{\varphi} =
       \tau^{2} h_{a b}(\varphi)
       \frac{d \varphi^{a}}{d \tau} \frac{d \varphi^{b}}{d \tau}},
       \eeq

is the integral of motion for eqs. (\ref{4.phi}).

 In (\ref{4.g})  $c_{i}$ are constants

 \beq{4.c_i} c_{i} = |\Sigma|^{\alpha^{i}}
  \exp{\Bigl[ \beta^{i}_{0}
  - \alpha^{i} \sum^{n}_{j=1}\beta^{j}_{0}d_{j} \Bigr]},
 \eeq
  $i = 1,\ldots, n$, obeying $\displaystyle{\prod_{i=1}^n c_i^{d_i}} =
  |\Sigma(\upsilon)|$.

 {\bf Flat $h$.} For the special case of a flat target space metric
                 $h_{a b}(\varphi) = h_{a b}$
  we get
    \beq{4.phi-t}
    \varphi^{a} = \alpha^{a}_{\varphi} \ln \tau  + \bar{\varphi}^{a}_0,
     \eeq
   where $\bar{\varphi}^{a}_0$ are constants, $a = 1, \dots, l$, and

     \beq{4.Ep}
       \mathcal{E}_{\varphi}= \frac{1}{2}
                h_{a b} \alpha^{a}_{\varphi} \alpha^{b}_{\varphi}.
       \eeq

 The scalar curvature (\ref{4.7}) reads in this case
      \beq{4.R}
       R[g] =  2 w \mathcal{E}_{\varphi} \tau^{-2}.
       \eeq

It diverges for $\tau \to + 0$ if $\mathcal{E}_{\varphi} \neq 0$.
Hence all solutions with $\mathcal{E}_{\varphi} \neq 0$ are
singular.

    Let $\mathcal{E}_{\varphi} = 0$.
    For Milne-type sets of parameters, i.e. when $d_i = 1$
  and $\alpha^{i} = 1$ for some $i$ ($\alpha^j = 0$ for
  all $j \neq i$)   the metric is regular, when
  either (\textbf{i}) $g^{(i)} = - w dy^i \otimes dy^i$, $M_i = \R$
  ($ - \infty < y^i < + \infty$),
  or (\textbf{ii}) $g^{(i)} = w dy^i \otimes dy^i$, $M_i$
  is a circle of length $L_i$ ($0 < y^i < L_i$)
  and $c_i L_i = 2 \pi$ (i.e. when the cone singularity is absent).

  For $\mathcal{E}_{\varphi} = 0$  the solutions with
  non-Milne-type   sets of the Kasner parameters are singular (at least)
  if  the Riemann tensor squared for metrics $g^i$ obeys: $R_{m_i n_i p_i q_i} R^{m_i n_i
  p_i  q_i} [g^i] \geq C_i$, for some $C_i$.
  (For $d_i = 1,2,3$  $R_{m_i n_i p_i q_i} [g^i] = 0$, e.g. due to  $R_{m_i n_i} [g^i] = 0$
  for $d_i = 2,3$.) This is valid since the
  Riemann tensor squared for the metric $g$ is divergent at $\tau = + 0$
  in this case,   see   \cite{IM-95a}.

  \textbf{Special  (steady state) solutions.}

 Now we consider the special case when  $\Sigma(v) =  0$.
 Due to (\ref{4.6}) we obtain
 \beq{4.6a}
  E_{\varphi} =  -\frac{1}{2} \sum^{n}_{i=1}d_{i}(v^i)^{2} \leq 0. \eeq

 We get in this case $\gamma_0 = \displaystyle{\sum^{n}_{i=1}d_{i}\beta^{i}_0} = {\rm
 const}$ and hence the scalar curvature

 \beq{4.7a} R[g] = 2 w E_{\varphi} e^{-2\gamma_{0}} \eeq

 and the volume scale factor $v = e^{\gamma_0}$ are constants.

 The ``synchronous'' variable is proportional to $u$
 ($\tau = e^{\gamma_{0}} u$).

Hence, we obtained a restriction for the energy $E_{\varphi}\leq
0$. For $E_{\varphi} = 0$ all $v^{i} = 0$ and we are led to a
static Ricci-flat solution.

This possibility occurs if the target space metric $h$ is not
positive-definite (e.g. there are phantom scalar fields for flat
$h$).  The solutions of such type with one (phantom) scalar field
were considered in  \cite{BZ0}. (They are called as steady state
solutions, see \cite{BZ0}.)

{\bf Solutions with acceleration.}
 Let $d_1 =3$ and $M_1 = \R^3$. The factor space $M_1$ may be considered as describing
 our space. In both cases there exist subclasses of solutions
 describing  accelerated expansion of our space.

 Indeed,
 for Kasner-like solutions with $w = -1$ one could make a replacement $\tau \mapsto \tau_0 - \tau$
 where $\tau_0$ is a constant (corresponding to the so-called ``big
 rip''). For such replacement the scale factor of $M_1$ reads

\beq{4.a} a_1(\tau)  =  c_{1} (\tau_0 - \tau)^{\alpha_1},  \eeq

$c_1 > 0$. For $\alpha_1 < 0$ we get an accelerated expansion of
``our space'' $M_1$.

Analogous consideration may be carried out for special (steady
state) solutions. For $w = -1$, $d_1 =3$ and $v^1 > 0$ we get an
accelerated expansion of 3-dimensional factor space $M_1$.

\subsection{Solutions with one curved Einstein
            space and $n-1$ Ricci flat spaces}

Here we put

 \beq{5.1} {\rm Ric}[g^{1}] = \xi_{1}g^{1}, \quad \xi_{1}
          \neq 0, \qquad {\rm Ric}[g^{i}] = 0,\quad i>1,
 \eeq

i.e. the first space $(M_{1},g^{1})$ is an Einstein space of non-zero
scalar curvature and other spaces $(M_{i}, g^{i})$ are Ricci-flat.

The  Lagrangian (\ref{3.10}) reads in this case

 \beq{5.L} L_{\beta} = \frac{1}{2} G_{ij}\dot{\beta^{i}}\dot{\beta^{j}}
 - \displaystyle{\frac{w}{2}\xi_{1}d_{1} \exp{(-2\beta^{1} +
 2\gamma_0)}}, \eeq

 where $ - \beta^{1} + \gamma_0 = u^{(1)}_i \beta^i$ and  $u^{(1)}_i =
- \delta^1_i  + d_i$.

The Lagrange equations corresponding to the Lagrangian (\ref{5.L})
are integrated in Appendix. The solution reads

 \bear{5.6a}
 \beta^{1} = \frac{1}{1 - d_{1}} \ln|f|
 + \upsilon^{1} u + \beta^{1}_{0}, \\
  \label{5.6b}
 \beta^{i} =   \upsilon^{i}u + \beta^{i}_{0},  \quad i > 1,
 \ear
 where $\beta^{i}_{0}$, $\upsilon^{i}$ are constants obeying

  \beq{5.6c} v^1   = \sum_{i=1}^n v^i d_i,
  \qquad  \beta_0^1 = \sum_{i=1}^n \beta_0^i d_i. \eeq

The function $f$ is following

 \bear{5.f} f = \left\{
      \begin{array}{ll}
        B \sinh(\sqrt{C}(u-u_{0})), \quad C> 0, \quad w\xi_{1}> 0; & \hbox{ } \\
        |\xi_{1}(d_{1} - 1)|^{1/2}(u-u_{0}),\quad C = 0, \quad w\xi_{1}>0; & \hbox{ } \\
        B  \sin(\sqrt{-C}(u-u_{0})), \quad C< 0, \quad w\xi_{1}>0; & \hbox{ } \\
        B \cosh(\sqrt{C}(u-u_{0})),\quad C> 0, \quad w\xi_{1}<0. & \hbox{ } \\
      \end{array}
    \right.
\ear

where $u_{0}$ and $C$ are constants and

\beq{5.7} B  = \sqrt{\frac{|\xi_{1}(d_{1} - 1)|}{|C|}}. \eeq

For  $\gamma_0$ we get

\beq{5.8} \gamma_0 = \beta^{1} - \ln{|f|}. \eeq

 The energy integral of motion $E_{\beta}$ corresponding to
 $L_{\beta}$ reads (see Appendix)

 \beq{5.8e} E_{\beta} = \frac{C d_1}{2 (1 - d_1)} + \frac 12 G_{ij} v^i v^j.
 \eeq

Using (\ref{5.6a}), (\ref{5.6b}) and (\ref{5.8}) we are led to the
relation for the metric

\bear{5.g} g = |f|^{\frac{2d_{1}}{1-d_{1}}}\exp{[2(\upsilon^{1} u+
\beta^{1}_{0})]}\left( w du \otimes du  + f^{2}g^{1}\right) +
\sum^{n}_{i=2}\exp{[2(\upsilon^{i} u + \beta^{i}_{0})]}g^{i}. \ear

 The ``scalar fields'' obey eqs. (\ref{3.5}) with the energy
constraint

\beq{5.8p} E_{\varphi} = \frac 12
  h_{a b}(\varphi)\dot{\varphi}^{a} \dot{\varphi}^{b} =
  \frac{C d_1}{2 (d_1 -1)} -\frac{1}{2}G_{ij}v^{i}v^{j}. \eeq

Here the constraints (\ref{5.6c}) on $\beta^{i}_{0}$,
$\upsilon^{i}$ should be kept in mind and the function $f$ is
defined in (\ref{5.f}).

Relations (\ref{5.6c}) are equivalent to the following ones

 \beq{5.8v} v^1   = \frac{1}{1 -d_1} \sum_{i=2}^n v^i d_i,
  \qquad  \beta_0^1 = \frac{1}{1 -d_1} \sum_{i=2}^n \beta_0^i d_i.
 \eeq

 Using  (\ref{5.6c}) and (\ref{5.8v})
 we exclude $v^1$ and obtain

 \beq{5.8vv}
 G_{ij}v^{i}v^{j} =  \sum_{i=1}^n d_i (v^i)^2
    - (\sum_{i=1}^n d_i v^i)^2=
  \frac{1}{d_1 -1}  (\sum_{i=2}^n d_i v^i)^2 +  \sum_{i=2}^n d_i (v^i)^2 \geq 0
 \eeq

and hence the relation (\ref{5.8p}) can be rewritten as follows

 \beq{5.8pp}
  \frac{1}{d_1 -1}  (\sum_{i=2}^n d_i v^i)^2 +  \sum_{i=2}^n d_i (v^i)^2
   =   \frac{C d_1}{(d_1 -1)} -  2 E_{\varphi} \geq 0. \eeq

In a special case of one (non-fantom) scalar field  ($h_{11} = 1$)
and $w= -1$ this solution was obtained earlier in \cite{BZ1,BZ2},
see also \cite{IM10}. The solutions with several scalar fields and
constant $h_{ab}$ were presented in \cite{I-03,I-05}. We note that
four-dimensional solutions with one scalar field (and
electromagnetic one) were obtained earlier in  \cite{KA-73}.

\section{Examples of geodesic solutions}

In this section we consider three  examples of solutions to
geodesic equations corresponding to the metric $h$ that may be
used for the cosmological type solutions above.

\subsection{Two-dimensional sphere.}

Let $h$ be a metric on a two-dimensional sphere $S^{2}$
   \beq{5.S2}
     h = d \theta \otimes d
    \theta  +   \sin^2 \theta   d \varphi \otimes d \varphi,
    \eeq

where $0 < \varphi < 2 \pi$ and $0 < \theta <  \pi$.
 The simplest solution to geodesic equations
(\ref{3.5}) for the metric reads

    \beq{5.S2g}
      \varphi = \omega u,   \qquad  \theta =0,
    \eeq

where $\omega $ is constant. Here  $E_{\varphi} = \frac12
\omega^2$.  The general solution to geodesic equations may be
obtained by a proper isometry $SO(3)$- transformation of the
solution from  (\ref{5.S2g}).

Here and in what follows   the first relation in (\ref{5.S2g})
should be understood as modulo $2 \pi$.



 \subsection{Two-dimensional de Sitter space.}

Now we put $h$ to be a metric on a two-dimensional de Sitter space
$dS^{2}$

   \beq{5.dS2}
     h =  - d \chi \otimes d \chi
      +   \cosh^2 \chi   d \varphi \otimes d \varphi,
    \eeq

where $0 < \varphi < 2 \pi$ and $ - \infty < \chi < + \infty$.
(For a review on the de Sitter space see \cite{MOS}.)

There are three basic solutions to geodesic equations (\ref{3.5})
in this case

    \bear{5.dS2g1}
      \varphi = \omega u,   \qquad  \chi =0, \\
      \label{5.dS2g2}
      \chi = v u, \qquad  \varphi = 0, \\
      \label{5.dS2g0}
     \tan{\varphi} =  \sinh \chi = mu,
     \ear

where $\omega, v$ and  $m $ are constants. For the energy we have
$E_{\varphi} = \frac12 \omega^2$, $- \frac12 v^2$ and $0$, for
 space-like, time-like and null geodesics, respectively. The
general solution to geodesic equations may be obtained by a proper
isometry $SO(1,2)$- transformation of the solutions from
 (\ref{5.dS2g1}) -  (\ref{5.dS2g0}).


 \subsection{The space with diagonal metric $h$.}

Here we consider a  diagonal metric

\beq{5.h} h = \varepsilon_{0}d\varphi \otimes d\varphi +
\sum^{l-1}_{k=1} \varepsilon_{k}A^{2}_{k}(\varphi)d\psi^{k}\otimes
d \psi^{k}, \eeq

where $\varepsilon_{0} = \pm 1$, $\varepsilon_{k} = \pm 1$ ($k >
0$) and all $A_{k}(\varphi) > 0$  are smooth functions.

The Lagrange function for the non-linear sigma model is given by

\beq{5.Lag} L_{\varphi} = \frac{1}{2} [
\varepsilon_{0}\dot{\varphi}^{2} +
\sum^{l-1}_{k=1}\varepsilon_{k}A^{2}_{k}(\phi)
(\dot{\psi}^{k})^{2}]. \eeq

Equations of motion for cyclic variables $\psi^{k}$

\beq{5.eq} \frac{d}{du}\left(\varepsilon_{k}
         A^{2}_{k}(\varphi)\dot{\psi}^{k}\right) = 0 \eeq

yield the following integrals of motion

\beq{5.C} \varepsilon_{k} A^{2}_{k}(\varphi)\dot{\psi}^{k} =
M_{k}, \eeq

$k = 1, \dots, l -1$.

Another constant of integration is the energy $E_{\varphi}$

\beq{5.E} E_{\varphi} = \frac{1}{2} [
 \varepsilon_{0}\dot{\varphi}^{2} + \sum^{l-1}_{k=1}\varepsilon_{k}
      A^{2}_{k}(\varphi)(\dot{\psi}^{k})^{2}] \eeq

which due to (\ref{5.C}) reads

\beq{5.EC} E_{\varphi} =
\frac{1}{2}[\varepsilon_{0}\dot{\varphi}^{2} +
\sum^{l-1}_{k=1}\varepsilon_{k} M^{2}_{k} A_k^{-2}(\varphi)]. \eeq

This relation implies the following quadrature

\beq{5.u} \displaystyle{\int_{\varphi_0}^{\varphi} \frac{d
 \bar{\varphi}}{\sqrt{2 \varepsilon_{0} E_{\varphi} -
 \sum^{l-1}_{k=1}\varepsilon_{0} \varepsilon_{k} M^{2}_{k}
  A_k^{-2}(\bar{\varphi})}}}= \pm (u - u_{0}), \eeq

which  implicitly defines the function $\varphi = \varphi (u)$.

Another quadratures just following from (\ref{5.C})

\beq{5.psi}
   \psi^{k} - \psi^k_0 = \displaystyle{\int_{u_0}^{u} d
\bar{u}  \varepsilon_{k} M_k A_k^{-2}(\bar{u})}, \eeq

which complete the integration of the geodesic equations for the
metric (\ref{5.h}).

For $A_{k}(\varphi) = \exp{(\lambda \varphi)}$, $\lambda \neq 0$,
the metric (\ref{5.h}) may describe either a part of de Sitter
space (if $\varepsilon_{0} = -1$, $\varepsilon_{k} = 1$, $k >0$)
or a part of anti-de Sitter space (if $\varepsilon_{1} = - 1$
$\varepsilon_{r} = 1$, $r \neq 1$). The case $l = 3$ is of
interest in  connection with the so-called AWE hypothesis
\cite{AWE} (based on Damour-Gibbons-Gudlach approach \cite{DGG})
aimed to describe the dark sector of the Universe.
In \cite{LLMO} a four-dimensional de Sitter sigma  model  coupled
 to  Einstein  gravity (e.g. describing  the  current
acceleration of the Universe) was studied.

\section{Spherically symmetric solutions}

In this section we study a subclass of spherically symmetric
solutions with the metric (\ref{5.g}) and ``scalar fields''
obeying (\ref{3.5}) and  (\ref{5.8p}). Here $g^{1}$ is a canonical
metric on the sphere $S^{d_{1}}$  and

\beq{6.1} w=1, \quad \xi_{1} = d_{1} -1, \eeq

where $d_{1} > 1$.

 Here we assume  $C > 0$. Then the governing function  looks like

\beq{6.2} f  = \frac{1}{\mu}\sinh{(\bar{\mu}u)}
 \eeq

where

 \beq{6.3} \bar{\mu} =\sqrt{C},\quad \mu  =
  \frac{\bar{\mu}}{d_{1} - 1} . \eeq

 For simplicity, we put $u_{0} = 0$ and  $\beta^{i}_{0} = 0$
 for all $i$.

By introducing a new radial variable $R = R(u)$

 \beq{6.4} \exp{[-2\bar{\mu} u]} = 1 - \frac{2\mu}{R^{d_{1}
                        -   1}} = F(R) = F, \eeq

with

 \beq{6.4r} R > R_0 \equiv  (2 \mu)^{1/(d_1 -1)} , \eeq

the solution for the metric can be rewritten as follows

\beq{6.5}
 g =  F^{b - 1} dR\otimes d R +
    R^{2}F^{b} g^{1}  + \sum^{n}_{i=2}F^{a_{i}}g^{i},
\eeq

where

\beq{6.6} b = \frac{1}{d_{1} - 1}-  \frac{v^{1}}{\bar{\mu}},
   \quad a_{i} = -
 \frac{v^{i}}{\bar{\mu}}, \quad 2\leq i\leq n. \eeq

Here we assign a metric for the time direction putting down $M_2=
 \R$ and $g^2 = -dt \otimes dt$.  Then  the metric (\ref{6.5}) reads

\beq{6.7} g = - F^{a_{2}} dt \otimes dt + F^{b - 1} dR\otimes d R
+ R^{2}F^{b} g^{1} + \sum^{n}_{i=3}F^{a_{i}}g^{i}, \eeq

where  due to (\ref{5.8v}), (\ref{5.8pp}) the constants $b$ and
$a_i$ satisfy the relations (see (\ref{6.3}) and (\ref{6.6}))

\beq{6.8}
 b = \frac{1}{d_{1} - 1}
 \left( 1 -  \sum^{n}_{i=2}d_{i}a^{i} \right)
 \eeq

 and

 \beq{6.9}
 \frac{1}{d_{1} - 1}\left(\sum^{n}_{i=2}d_{i}a_{i}\right)^{2} +
 \sum^{n}_{i=2}d_{i}a^{2}_{i} = \frac{d_{1}}{d_{1} -1} -
 e_{\varphi}. \eeq

Here we denote
\beq{6.10}
 e_{\varphi} \equiv
 \frac{2E_{\varphi}}{C}
 = \frac{h_{ab}(\varphi)\dot{\varphi}^{a}\dot{\varphi}^{b}}{C} =
 \frac{1}{\bar{\mu}^{2}}h_{ab}(\varphi)
 \frac{d\varphi^{a}}{dR}\frac{d\varphi^{b}}{dR}R^{2d_{1}}F^{2}.
\eeq

The ``scalar fields'' $\varphi^a = \varphi^a (R)$ obey the
equations of motions
 \beq{6.10eq}
  -  \frac{d}{dR} \left( F(R)  R^{d_{1}} h_{b c}(\varphi)
  \frac{d\varphi^{b}}{dR}\right) +
 \frac{1}{2} F(R)  R^{d_{1}} h_{ab, c}(\varphi)
  \frac{d\varphi^{a}}{dR} \frac{d\varphi^{b}}{dR} = 0,
\eeq
 $c = 1,\ldots, l$, which are equivalent to eqs. (\ref{3.5})
 (see (\ref{6.4})).

 These equations are nothing more than Euler-Lagrange equations
 for the action
  \beq{6.10S}
     S_{\varphi} = - \frac{1}{2} \int dR
     F(R)  R^{d_{1}} h_{ab}(\varphi) \frac{d\varphi^{a}}{dR} \frac{d\varphi^{b}}{dR}
     = \frac{1}{2} \int du h_{ab}(\varphi) \frac{d\varphi^{a}}{du}
     \frac{d\varphi^{b}}{du}.
  \eeq

 Thus we have obtained a family of spherically symmetric solutions to field
 equations with a sigma model source which are given by relations (\ref{6.7}) -
 (\ref{6.10}). These solutions are defined up to solutions to
 ''scalar fields equations'' (\ref{6.10eq}) which are in fact the
 geodesic equations for the metric $h$ (\ref{3.5}) rewritten in
 the $R$-variable.

 {\bf Example: the case of constant $h_{ab}.$} Let us consider a
 special case when $h_{a b}(\varphi) = h_{a b}$ are constants.
  It follows from (\ref{3.15}) and  (\ref{6.4}) that

\beq{6.10F} \varphi^{a} = \frac{1}{2} q^{a} \ln F(R) +
\varphi^{a}_0, \eeq

where $q^a = - \upsilon^{a}_{\varphi}/ \bar{\mu} $  and
$\varphi^{a}_0$ are integration constants, $a = 1, \dots, l$. For
the energy parameter we get from  (\ref{4.12})
 \beq{6.10e}
  e_{\varphi} =  h_{ab} q^{a} q^{b}.
 \eeq

Due to Hilbert-Einstein  field equations (\ref{1.4E}) (with
$T_{MN}$ from (\ref{1.4F})) we get the following relation for the
scalar curvature

 \beq{6.11E}
  R[g] = g^{MN}  R_{MN} = g^{MN}  T_{MN}/ (1 - D/2) =
  g^{MN}  h_{ab}(\varphi) \partial_{M}\varphi^a  \partial_{N}
  \varphi^{b}.
 \eeq

  which implies (for the solution under consideration)

 \beq{6.11R}
  R[g] =
  F^{1-b} h_{ab}(\varphi) \frac{d\varphi^{a}}{dR} \frac{d\varphi^{b}}{dR}
  = e_{\varphi} \bar{\mu}^{2}R^{-2d_{1}}F^{-1-b}.
  \eeq

   Due to (\ref{6.11E}) the Hilbert-Einstein equations (\ref{1.4E})
   can be rewritten in an equivalent form

   \beq{6.11ER}
   R_{MN} =   h_{ab}(\varphi) \partial_{M}\varphi^a  \partial_{N}
                        \varphi^{b}.
    \eeq

  In what follows we denote
   \beq{6.11bb}
    b_{0} = \displaystyle{\sum^{n}_{i=2}d_{i}a_{i}}.
   \eeq

 We remind that  $e_{\varphi}$ is defined by  (\ref{6.10}) and obeys the
inequality following from (\ref{6.9})
 \beq{6.11e}
   e_{\varphi} \leq \frac{d_{1}}{d_{1} -1}.
 \eeq

 \textbf{Proposition 1.}
 {\it For all $a_i$ and   $e_{\varphi}$
  obeying (\ref{6.9})
  \beq{6.12}
   |b_{0}(a)| \leq
  \sqrt{\frac{(d_{1} - 1)(D-d_{1} -1)}{D-2}}\sqrt{\frac{d_{1}}{d_{1}
  -1} - e_{\varphi}}. \eeq
  Maximum and minimum  of the function $b_{0}(a)$ are attained at
  the points $a_{+}$ and $a_{-}$, respectively,
  where $a_{\pm} = (a_{\pm,i})$
  \beq{6.12mm}
   a_{\pm,i} = \pm \sqrt{\frac{(d_{1} - 1)}{(D-2)(D-d_{1} -
  1)}}\sqrt{\frac{d_{1}}{d_{1} -1} - e_{\varphi}}, \eeq
  for $i = 2, \ldots,n$. } \\

\textit{Proof.} For $e_{\varphi} = \frac{d_{1}}{d_{1} -1}$
 the proposition is trivial. Let us consider the
 case $ A = \frac{d_{1}}{d_{1} -1} - e_{\varphi} > 0$.
 We prove the proposition by the method of
Lagrange multipliers. We introduce a new variable $\lambda$ called
a Lagrange multiplier, and study the Lagrange function defined by

 \beq{6.13}
 \bar{b}_{0}(a,\lambda) = b_{0}(a) - \lambda(Q(a)-A), \eeq

 where

 \beq{6.13Q}
  Q(a)= \frac{1}{d_{1} -
  1}\left(\sum^{n}_{i=2}d_{i}a_{i}\right)^{2} +
 \sum^{n}_{i=2}d_{i}a^{2}_{i}. \eeq

 For the points of extremum we get

\bear{6.14a} \frac{\partial \bar{b}_{0}}{\partial \lambda} = 0
 \quad \Leftrightarrow \quad Q = A,
 \\ \label{6.14b}
 \frac{\partial \bar{b}_{0}}{\partial a_{i}} = 0 \quad \Leftrightarrow
\quad 1- 2\lambda\left[\frac{1}{d_{1} - 1}\sum^{n}_{j=2}d_{j}a_{j}
 + a_{i}\right] = 0, \ear
 $i = 2, \ldots,n$

  It follows from (\ref{6.14b}) that all $a_i$
  are coinciding

 \beq{6.15a}
 a_{i} =  \frac{(d_{1} - 1)}{D-2}\frac{1}{2\lambda}
 \eeq
 for $i = 2, \ldots,n$. The substitution of (\ref{6.15a})
 into relation (\ref{6.14a}) gives us two points of extremum
 of the function $b_{0}(a)$ on the  $(n-2)$-dimensional ellipsoid
 $Q(a) = A$ which are given by relations (\ref{6.12mm}).
 The point $a_{+}$ is the point of maximum and $a_{-}$ is the
 point minimum. We get the  inequality
  $b_0(a_{-})  \leq b_0(a) \leq b_0(a_{+})$
  coinciding with (\ref{6.12}). The proposition is proved.

\textbf{Proposition 2.} {\it Let $e_{\varphi}\neq 0$ and

\beq{6.16a}
  -\frac{d_{1}(D-1)}{D - d_{1} - 1} <
  e_{\varphi} \leq \frac{d_{1}}{d_{1} - 1}.
 \eeq

 Then the scalar curvature $R[g]$ for the metric (\ref{6.5})
 with parameters $b$, $a_i$ obeying (\ref{6.8}) and (\ref{6.9})
 is divergent at $R = R_{0} + 0$, i.e.
  \beq{6.16b}
 R[g] \rightarrow + \infty \quad \textrm{for} \quad R \rightarrow R_{0}.
 \eeq
}

\textit{Proof.} It follows from the Proposition 2
 and the first inequality for the energy parameter $e_{\varphi}$  in (\ref{6.16a})
 that
 \beq{6.17a}
 |b_{0}(a)| < d_{1},
 \eeq
 and hence ($d_{1} > 1$)
 \beq{6.17b}
  - 1 - b = \frac{1}{d_{1} - 1}(-d_{1} + b_{0}) <0.
 \eeq
 Using the relation for the scalar curvature (\ref{6.11R}),
 the inequality (\ref{6.17b}) and $e_{\varphi} \neq 0$ we get
 the divergence of $R[g]$ for $R \to R_{0}$. This completes
 the proof of the proposition.

 Now we consider the special case $e_{\varphi} = 0$. In this case
 the stress-energy tensor for $\varphi$ vanishes and the metric
 $g$ is Ricci-flat, i.e. $R_{MN}[g]$ = 0. Thus we are led to
 a vacuum solution from \cite{FIM-91} which was considered in \cite{IM-95a}.
 Recently the solution from \cite{FIM-91} was intensively studied
 in \cite{JPL}.

 In what
 follows we are interested in the case of non-zero gravitational
 mass, i.e. we put $a_2 \neq 0$.

  \textbf{Proposition 3.} {\it Let $g$ be a  metric (\ref{6.7})
   with the parameters $b$, $a_i$
  obeying (\ref{6.8}), (\ref{6.9}),  $e_{\varphi} = 0$ and $a_2 \neq 0$.
  Let the Riemann tensor squared  for any metric $g^i$, $i =
  3,...,n$, obey a self-boundedness   condition
    \beq{6.18i}
    I_i = R_{m_i n_i p_i q_i} R^{m_i n_i
  p_i  q_i} [g^i]  \geq C_i,
    \eeq
  for some $C_i$.
  Then for all sets $a = (a_2,...,a_n) \neq (1,0,...,0)$ the
  Riemann tensor squared  for the metric (\ref{6.7})
  is divergent at $R = R_{0} + 0$, i.e.
    \beq{6.18}
    R_{MNPQ} R^{MNPQ}[g] \rightarrow
     + \infty \quad \textrm{for} \quad R \rightarrow R_{0}.
    \eeq
 }

 The Proposition 3 is a special case of the Proposition 6 from
 \cite{IM-95a}. The  Propositions 2 and 3 may be summarized as
 the following proposition. We note that the condition
 (\ref{6.18i}) is satisfied  for any metric $g^i$ of Euclidean
 signature since $I_i  \geq 0$ in this case.
 It is valid also for any $g^i$  of dimension $d_i = 1$
  and for any Ricci-flat $g^i$ of dimension $d_i = 2, 3$ since in all
  these cases the Riemann tensor of $g^i$ is zero and hence $I_i =0$.

  \textbf{Proposition 4.} {\it Let $g$ be a  metric (\ref{6.7})
   with the parameters $b$, $a_i$
  obeying (\ref{6.8}), (\ref{6.9}),  $a_2 \neq 0$ and
  (\ref{6.16a}). Let the Riemann tensor squared  for any metric $g^i$, $i =
  3,...,n$, obey a self-boundedness   condition (\ref{6.18i}).
   Then the regular horizon of the metric (\ref{6.7}) at $R = R_0$
   takes place if and only if
    \beq{6.19b}
     a_2 - 1 = a_3 = ... = a_n =  e_{\varphi} = 0.
    \eeq
 }

   The Proposition 4 may be considered as a restricted version of the
   ``no-hair'' theorem for the metric (\ref{6.7}). For the case
   of the  positive-definite sigma model metric  $h_{ab}(\varphi)$
   (i.e. when $e_{\varphi} \geq 0$) it was proved for more
   general assumptions in \cite{BFM}  (see Theorem 5 therein).

     For the set of parameters from (\ref{6.19b}) we get

    \beq{6.20} g = g_T   + \sum^{n}_{i=3} g^{i}, \eeq

    where  $g_T = - F dt \otimes dt +  F^{-1} dR \otimes d R   + R^{2} g^1$
    is the metric of the $(2+d_1)$-dimensional Tangherlini black hole
    solution \cite{Tan}.

    {\bf Remark.} Let all $h_{ab}$ be constant. Then the relation
     $e_{\varphi} = 0$ reads as $h_{ab} q^{a} q^{b} = 0$, where
     the scalar charges $q^{a}$ appeared previously for the  scalar field
     solutions (\ref{6.10F}). For the positive- or negative-definite
     matrix $h_{ab}$ we get $q^a = 0$ for all $a$.
     If the matrix $(h_{ab})$ is a semi-definite one, i.e. it has a signature
     $(-,...,-, +, ..., +)$ then there exist solutions  $\varphi^{a}$
     (\ref{6.10F}) with two or more non-zero $q^{a}$. However these
     solutions are singular on a horizon. The only regular
     solutions are trivial ones $\varphi^{a} = \varphi^{a}_0$.
     So, we do not obtain   non-trivial ``scalar hairs'' in the
     case of constant $h_{ab}$. Analogous situations takes place for the   $dS^2$ sigma
     model solution with $e_{\varphi} = 0$ (see (\ref{5.dS2g0})):
     $\varphi = \arctan( \frac{1}{2} q \ln F(R))$,
     $\chi = {\rm arcsinh}( \frac{1}{2} q \ln F(R))$ with $q = -
     m/\bar{\mu}  \neq 0$. These solutions are also singular
     at the horizon.

   {\bf The PPN parameters for $d_1 = 2$.} Let us consider the
    case $d_1 = 2$. The pure gravitational solution (without scalar fields)
    was obtained in cite \cite{BrIvMel-89} and generalized to the
    scalar-vacuum case (with one scalar field) in
    \cite{Ivashchuk-PhD-89},
    see also \cite{I-10}. The calculations of PPN parameters
    for a 4-dimensional part of the metric (\ref{6.7}) with $a_{2}
    \neq 0$

    \beq{6.21} g^{(4)} = - F^{a_{2}} dt \otimes dt + F^{b - 1} dR\otimes d R
      + R^{2}F^{b} g^{1} , \eeq

    and $F = 1 - \frac{2\mu}{R}$ give us the following
    relations \cite{BIM-92,DPM-92}
    (see also  \cite{EMCZ-11})
    \beq{6.22}
      \beta = 1, \qquad \gamma = 1 +
      \frac{1}{a_2} \displaystyle{\sum^{n}_{i=3}d_{i}a_{i}}.
      \eeq

     For
         \beq{6.23}
         \displaystyle{\sum^{n}_{i=3}d_{i}a_{i}} = 0
        \eeq

     and $a_2 \neq 0$ we are led to a subclass of solutions with $\beta = \gamma = 1$
     for the 4-dimensional metric (\ref{6.21}). The solution of
     such type were called in \cite{EMCZ-11} (for pure gravity) as latent
     solitons \cite{EMCZ-11}. The four-dimensional section of the
     latent soliton metric gives the same PPN parameters
     $\beta = 1$ and $\gamma = 1$ as  the Schwarzschildian metric does,
     i.e. gravitational experiments lead to the same results for
       $g^{(4)}$ and for the metric of the Schwarzschild solution.

  \section{Conclusions}

 In this paper  we have  considered a $D$-dimensional model of gravity with
 non-linear ``scalar fields'' governed by a sigma model action (as a matter source).
 The model is defined on the product  manifold $M$, which contains $n$ Einstein
 factor spaces $M_1, \dots M_n$.

We have  obtained general cosmological type solutions
corresponding to the field equations in two cases: when either all
factor spaces are Ricci-flat or when  only one factor space,
$M_1$, has nonzero scalar curvature. The solutions are defined up
to solutions of geodesic equations corresponding to the sigma
model target space. We have considered several examples of sigma
models, e.g. with $S^2$ and $dS^2$ target spaces. It is shown that
for certain parameters cosmological solutions may describe an
accelerated expansion of  3-dimensional factor space.

Here we have also  studied a subclass of spherically symmetric
solutions with $\sinh$-behaviour of the governing function $f$
from (\ref{5.f}). We have proved a restricted version of the
``no-hair theorem'' (Proposition 4) when the scalar energy
parameter $e_{\varphi}$ obeys restriction  (\ref{6.16a}) and all
factor space metrics $g^i$, $i =3, \dots, n$, have Euclidean
signatures. We have found for $d_1 = 2$  a subclass of latent
solitonic solutions generalizing those from ref. \cite{EMCZ-11}.

An open problem here is to generalize  the ``non-hair'' theorem
for the case of all  types of the governing function $f$ and all
values of the energy parameter $e_{\varphi}$ as it was done
recently for   $D=4$ case (with one scalar field) in \cite{BFZ}.

\begin{center}
 {\bf Acknowledgments}
 \end{center}

This work was supported in part by the FTsP ``Nauchnie i
nauchno-pedagogicheskie kadry innovatsionnoy Rossii'' for the
years 2009-2013.

 \renewcommand{\theequation}{\Alph{subsection}.\arabic{equation}}
 \renewcommand{\thesection}{}
 \renewcommand{\thesubsection}{\Alph{subsection}}
 \setcounter{section}{0}

 \section{Appendix}

\subsection{Solutions governed by Liouville equation}

\renewcommand{\theequation}{\Alph{section}.\arabic{equation}}

Here we consider a  Toda-like system with the following Lagrangian

\beq{A.1} L = \frac 12 \langle \dot{\beta}, \dot{\beta}\rangle -
A\exp{(2\langle b, \beta \rangle)}, \eeq

where $\beta \in \mathbb{R}^{n}$, $A \neq 0$, $b \in
\mathbb{R}^{n}$. Here (in Appendix) $\dot{\beta} = \frac{d
\beta}{dt}$. The scalar product for vectors belonging to $\R^{n}$
is defined by

\beq{A.2} \langle v_1, v_2 \rangle = G_{ij} v_1^{i}
 v_2^{j}, \eeq

where $(G_{ij})$ is a non-degenerate symmetric matrix (e.g. given
by (\ref{3.3})).

 The Lagrange equations corresponding to the model (\ref{A.1}) read
 (in a condensed vector form)

\beq{A.3}  \ddot{\beta} + 2A b \exp{(2\langle b, \beta \rangle)}
  =  0 .   \eeq

Let $\langle b,b \rangle \neq 0$.

 Eqs. (\ref{A.3}) is exactly integrable and the solution may be
written

\beq{A.4} \beta = \frac{b}{\langle b,b \rangle} q + v t + \beta_0,
\eeq

where $\langle b, b\rangle \neq0$ and $v,\beta_0 \in
\mathbb{R}^{n}$  are constant vectors obeying

\beq{A.5} \langle v, b\rangle = \langle \beta_0,b \rangle = 0.
\eeq

The function $q=q(t)$ obeys the Liouville equation

\beq{A.6} \ddot{q} + 2A \langle b,b \rangle e^{2q} = 0. \eeq

The solution to Liouville equation reads

\beq{A.7} q = - {\rm ln}|f|, \eeq

where

\bear{A.L} f = \left\{
      \begin{array}{ll}
        B \sinh(\sqrt{C}(t-t_{0})), \quad C> 0, \quad \bar{A}< 0 & \hbox{;} \\
        |2\bar{A}|^{1/2}(t-t_{0}),\quad C = 0, \quad \bar{A}<0 & \hbox{;} \\
        B \sin(\sqrt{-C}(t-t_{0})), \quad C< 0, \quad \bar{A}<0 & \hbox{;} \\
        |B|\cosh(\sqrt{C}(t-t_{0})),\quad C> 0, \quad \bar{A}>0 & \hbox{;} \\
      \end{array}
    \right.
\ear

here we put $\bar{A} = A \langle b,b\rangle$ and

\beq{A.R} B = \sqrt{\frac{2|A\langle b,b \rangle|}{|C|}}. \eeq

The energy function corresponding to the Lagrangian (\ref{A.1})
reads

\beq{A.8} E = \frac{1}{2}\langle \dot{\beta},\dot{\beta} \rangle +
            A e^{2 \langle b, \beta \rangle}. \eeq

After substitution of (\ref{A.4}) to (\ref{A.8}) we obtain

\beq{A.9} E = E_{T} + \frac 12 \langle v, v \rangle, \eeq
  where

\beq{A.10} E_{T} = \frac{1}{2\langle b,b\rangle} \dot{q}^{2} +
                    Ae^{2q}. \eeq

Due to (\ref{A.7}) and (\ref{A.L}) we get
 \beq{A.11}
  E_{T} = \frac{C}{2 \langle b,b \rangle} .
   \eeq

\textbf{Proposition 1A.}
  For $\langle b,b \rangle \neq 0$ all solutions to  Lagrange equations (\ref{A.3}) are
  covered by the relations (\ref{A.4}), (\ref{A.5}), (\ref{A.7}) and (\ref{A.L}).\\

\textbf{Proof.}\\

 $\Rightarrow$. It is obvious that the solutions (\ref{A.4}), (\ref{A.5}) with $q$ from
 (\ref{A.7}), (\ref{A.L})
 obey the equations of motion (\ref{A.3}).\\

 $\Leftarrow$. Let us show that the relations (\ref{A.4}), (\ref{A.5}), (\ref{A.7}) and (\ref{A.L})
 followed from (\ref{A.3}).\\

Let $q = \langle b, \beta \rangle$ , $y = \beta - (b q)/ \langle
b,b \rangle$. It is obvious that $\langle b, y \rangle = 0$. It
follows from  (\ref{A.3}) that the equation (\ref{A.6}) is
satisfied identically and

 \bear{A.13} \ddot{y} = 0  \quad \Rightarrow \quad y= v t +
 \beta_0, \ear

 where constant vectors $v$ and $\beta_0$ obey (due to $\langle b, y \rangle = 0$)

 \beq{A.14}
  \langle b, v \rangle = \langle b, \beta_0 \rangle
  =0.
 \eeq

  Hence
 \beq{A.15}
 \beta = \frac{bq}{\langle b,b\rangle} + y =  \frac{bq}{\langle b,b\rangle} + v t +
 \beta_0,
  \eeq
 where $q = q(t)$ obeys (\ref{A.6}) and hence it is given by
 relations (\ref{A.7}) and (\ref{A.L}).

 The Proposition 1A is proved.

Let us introduce a dual vector $u = (u_i)$: $u_i = G_{ij}b^j$.
Then we get $u(\beta) = u_i \beta^i = \langle b, \beta \rangle$,
$(u,u) = G^{ij} u_i u_j = \langle b, b \rangle $, where $(G^{ij})
= (G_{ij})^{-1}$, and the solution (\ref{A.4}) reads

\beq{A.16} \beta^i = - \frac{u^i}{(u,u)} {\rm ln}|f| + v^i t +
           \beta_0^i, \eeq

$i = 1, \dots, n$,  where $(u,u) \neq0$,

 \beq{A.17} u(v) = u_i v^i =0, \qquad u(\beta_0) = u_i \beta_0^i
            =0, \eeq

and the function  $f$ is defined in (\ref{A.L}) with

\beq{A.Ru} B = \sqrt{\frac{2|A (u,u)|}{|C|}}, \qquad \bar{A} = A
           (u,u). \eeq

For the energy (\ref{A.10}) we obtain from (\ref{A.9}) and
(\ref{A.11})

 \beq{A.18} E = \frac{C}{2 (u,u)} + \frac 12 G_{ij} v^i v^j. \eeq

 {\bf Example.} Let us consider a Lagrange system from Section 3
  with parameters: $A = \frac{w}{2} \xi_{1}d_{1}$,
  $u_i = u^{(1)}_i =  - \delta^1_i  + d_i$, $d_1
  > 1$.  Then due to (\ref{3.3u}) and (\ref{3.13}) we
  get   $u^i = - \frac{\delta^{1i}}{d_1}$ and $(u,u)= \frac{1}{d_1}
  -1 < 0$.  The solution  reads

  \beq{A.16e} \beta^i =  \frac{\delta^i_1}{1 - d_1} {\rm ln}|f| + v^i t +
               \beta_0^i, \eeq

 $i = 1, \dots, n$, with constraints

  \beq{A.17e} v^1 = \sum_{i=1}^n v^i d_i,
  \qquad  \beta_0^1 = \sum_{i=1}^n \beta_0^i d_i \eeq

  imposed.
  In (\ref{A.L}) we should put
  $\bar{A} =  \frac{w}{2} \xi_{1} (1 -d_{1})$ and
  $B = \sqrt{\frac{|\xi_{1}| (d_1 - 1)}{|C|}}$.

  For the energy we get
 \beq{A.18e} E = \frac{C d_1}{2 (1 - d_1)} + \frac 12 G_{ij} v^i v^j.
 \eeq

\end{document}